\documentclass[aps,prl,reprint,superscriptaddress,amsmath,amssymb]{revtex4-2}
\usepackage{bm}
\usepackage{graphicx}
\usepackage[colorlinks=true,citecolor=blue,urlcolor=blue,linkcolor=blue]{hyperref}

\newcommand{\dd}{\mathrm{d}}
\newcommand{\wmax}{\omega_{\rm max}}

\begin{document}

\title{Analytical Singular-Value Structure of Analytic-Continuation Kernels\\
from Slepian Information Theory}

\author{Masayuki Ohzeki}
\email{mohzeki@tohoku.ac.jp}
\affiliation{Graduate School of Information Sciences, Tohoku University, Sendai 980-8579, Japan}
\affiliation{Department of Physics, Institute of Science Tokyo, Tokyo 152-8551, Japan}
\affiliation{
Research and Education Institute for Semiconductors and Informatics, Kumamoto University, Kumamoto 860-8555, Japan}
\affiliation{Sigma-i Co., Ltd., Tokyo 108-0075, Japan}

\date{\today}

\begin{abstract}
Analytic continuation from imaginary-time Green's functions to real-frequency spectra is a central ill-posed inverse problem in quantum many-body physics.  We show that the thermal kernel admits an analytical generalized singular-value structure once its purely dynamical part is separated from the statistical weight imposed by the heat bath.  The dynamical kernel is the imaginary-bandwidth continuation of Slepian's finite Fourier transform and is governed by the same Sturm-Liouville algebra that yields prolate spheroidal wave functions.  Fermionic and bosonic statistics then enter as gauge transformations of the frequency-space inner product, producing self-adjoint effective potentials but no numerical kernel diagonalization.  The Shannon number, $N_c=\beta\wmax/\pi$, fixes the upper information capacity of this pure Laplace channel.  Finally, the optimal sampling points are obtained as eigenvalues of a Legendre colleague matrix, giving a deterministic compressed-sensing grid without iterative root searches.
\end{abstract}

\maketitle

{\it Introduction.---}
Imaginary-time quantum Monte Carlo (QMC) methods provide statistically controlled access to thermal correlation functions, but experimentally relevant spectra live on the real-frequency axis.  Their relation is a Fredholm equation of the first kind,
\begin{equation}
G^\nu(\tau)=\int_{-\wmax}^{\wmax}K^\nu(\tau,\omega)A(\omega)\,\dd\omega ,
\label{eq:fredholm}
\end{equation}
where $\nu=F,B$ labels fermionic or bosonic statistics.  Inverting Eq.~(\ref{eq:fredholm}) is the analytic-continuation problem.  It is notoriously unstable: singular values of the kernel collapse exponentially, so statistical noise in $G^\nu$ is amplified into uncontrolled spectral structure \cite{JarrellGubernatis1996}.  The standard responses to this instability have been to impose prior structure: rational Pad\'e continuation \cite{VidbergSerene1977}, maximum entropy and Bayesian inference \cite{SilverSiviaGubernatis1990,JarrellGubernatis1996}, and stochastic analytic continuation or optimization \cite{Sandvik1998,ShaoSandvik2023}.  These methods remain indispensable, but their output depends on model selection, entropy or temperature schedules, positivity and sum-rule constraints, and diagnostics for distinguishing physical sharp features from amplified noise.

The introduction of sparse modeling changed the numerical viewpoint: instead of reconstructing all singular directions, one identifies the components that survive the noise level by an $L_1$-regularized inverse problem \cite{Otsuki2017,Yoshimi2019}.  In parallel, the intermediate representation (IR) exposed the low-rank structure of finite-temperature Green's functions and enabled sparse sampling in imaginary time and Matsubara frequency \cite{Shinaoka2017,Li2020,Shinaoka2020}.  This program has since expanded into discrete Lehmann representations \cite{Kaye2022}, hybrid sparse-modeling--Pad\'e strategies \cite{Motoyama2022}, causality-preserving Nevanlinna continuations \cite{FeiYehGull2021}, machine-learned optimizers \cite{HuangYang2022}, and integrated software ecosystems such as ACFlow \cite{HuangACFlow2026}.  A recent independent work has further identified the finite-Laplace/oblate-spheroidal origin of compression in the standard IR kernel \cite{Misawa2026}.  Yet the corresponding optimal measurement grid and its relation to sparse inverse solvers have remained open.

This Letter gives a closed analytical route to that algorithmic step.  We factor the thermal kernel into a universal dynamical transmission factor and a statistics-dependent metric.  The former is mapped to Slepian's information theory of time- and band-limited signals; the latter becomes a gauge transformation of the Hilbert-space measure.  The result is an exact generalized SVD of the fermionic and bosonic kernels in the weighted metric, together with an algebraic construction of the corresponding optimal sampling grid.

{\it Slepian's finite Fourier transform.---}
Before turning to Green's functions, it is useful to recall the classical problem solved by Slepian, Landau, and Pollak \cite{SlepianPollak1961,LandauPollak1961,LandauPollak1962,Slepian1978}.  Let a signal be observed only in a finite time interval $x\in[-1,1]$ and be band-limited to $y\in[-1,1]$.  The finite Fourier transform
\begin{equation}
({\cal F}_c f)(x)=\int_{-1}^{1}e^{icxy}f(y)\,\dd y
\label{eq:slepian-fourier}
\end{equation}
is not diagonal in plane waves because both time and frequency have been truncated.  Slepian's key observation was that ${\cal F}_c$ commutes with a local Sturm-Liouville operator,
\begin{equation}
{\cal D}^{(F)}_z=\frac{\partial}{\partial z}
\left[(1-z^2)\frac{\partial}{\partial z}\right]-c^2z^2 ,
\qquad
[{\cal F}_c,{\cal D}^{(F)}]=0 .
\label{eq:slepian-reference}
\end{equation}
Thus the singular functions of the nonlocal integral transform are the prolate spheroidal wave functions (PSWFs), obtained from a stable differential equation rather than from a brute-force SVD.  The same theory gives the Shannon number $N_c=2c/\pi$, the number of independent degrees of freedom that can pass through a finite time-bandwidth window.

This classical construction is the reference frame for the present work.  The Green-function kernel differs from Eq.~(\ref{eq:slepian-fourier}) in two controlled ways: the unitary Fourier phase $e^{icxy}$ is replaced by the imaginary-time Laplace factor $e^{-cxy}$, and fermionic or bosonic statistics multiply this universal factor by thermal weights.  We now show that the first change is an analytic continuation of the Slepian algebra, while the second is a gauge transformation of the frequency-space metric.  Details of the kernel symmetrization and Slepian mapping are given in Secs.~S1 and S2 of the Supplemental Material.

{\it Slepian mapping and Shannon limit.---}
Introduce
\begin{equation}
x=\frac{2\tau}{\beta}-1,\qquad
y=\frac{\omega}{\wmax},\qquad
c=\frac{\beta\wmax}{2}.
\end{equation}
The physical kernels on $[-1,1]^2$ separate as
\begin{align}
K^F(x,y)&=\frac{1}{2}\frac{e^{-cxy}}{\cosh(cy)},\label{eq:kfermion}\\
K^B(x,y)&=\frac{\wmax}{2}\frac{y e^{-cxy}}{\sinh(cy)}. \label{eq:kboson}
\end{align}
All dependence on quantum statistics is confined to the $y$-space weights.  The common factor
\begin{equation}
{\cal K}_c(x,y)=e^{-cxy}
\end{equation}
is the pure imaginary-time phase-transfer channel.  It satisfies
\begin{equation}
{\cal D}^{(L)}_x{\cal K}_c(x,y)={\cal D}^{(L)}_y{\cal K}_c(x,y),
\end{equation}
where
\begin{equation}
{\cal D}^{(L)}_z=\frac{\partial}{\partial z}
\left[(1-z^2)\frac{\partial}{\partial z}\right]+c^2z^2.
\label{eq:commutation}
\end{equation}
This is the imaginary-bandwidth continuation of the Slepian Sturm-Liouville operator.  Hence ${\cal K}_c$ has the bilinear expansion
\begin{equation}
e^{-cxy}=\sum_{n=0}^{\infty}\mu_n\,\psi_n(x;c)\psi_n(y;c),
\label{eq:slepian-expansion}
\end{equation}
where $\psi_n$ are PSWFs analytically continued in the bandwidth parameter, equivalently oblate spheroidal functions.  The information capacity of this pure channel is therefore set by the Slepian-Landau-Pollak counting law,
\begin{equation}
N_c=\frac{2c}{\pi}=\frac{\beta\wmax}{\pi}.
\label{eq:shannon}
\end{equation}
Components with $n\gg N_c$ encode the exponentially small tail of the channel and are therefore the modes most vulnerable to numerical and QMC noise.

The Laplace version of this statement is less familiar than Slepian's original Fourier problem because the kernel is no longer unitary.  The present formulation makes the point directly visible: the exposed eigenvalues $\mu_n$ of Eq.~(\ref{eq:slepian-expansion}) can be evaluated from the PSWF coefficients without any SVD of a discretized kernel, as detailed in Sec.~S4 of the Supplemental Material.  Figure~\ref{fig:eigenvalue-cliff} shows the ordered magnitudes for the same parameter used below, $c=20$ ($\beta\wmax=40$).  The sign alternation of $\mu_n$ is absorbed into the right singular-vector convention; the generalized singular values of the fermionic and bosonic kernels are obtained from $|\mu_n|$ by the statistical prefactors in Eq.~(\ref{eq:svd-pieces}).  Around the Shannon index $N_c=12.73$, the spectrum has already fallen by more than seven orders of magnitude and then rapidly enters the exponentially small tail.

\begin{figure}[t]
\includegraphics[width=\columnwidth]{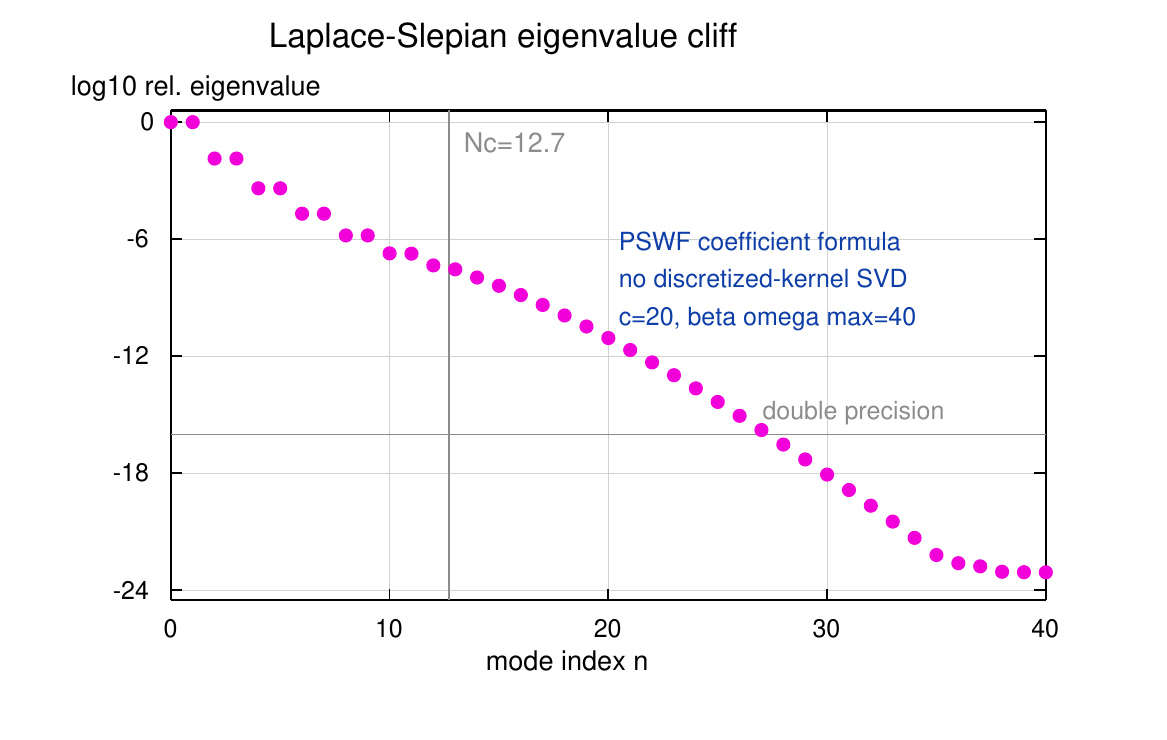}
\caption{Eigenvalue cliff of the pure Laplace-Slepian kernel $\mathcal K_c(x,y)=e^{-cxy}$ for $c=20$.  The plotted quantity is the ordered magnitude $|\mu_n|/|\mu_0|$ obtained from the Legendre coefficients of the analytically continued PSWFs, not from a numerical SVD of the kernel.  The vertical line marks the Shannon number $N_c=2c/\pi=12.73$, and the horizontal line indicates double-precision scale.}
\label{fig:eigenvalue-cliff}
\end{figure}

{\it Gauge-transformed analytical SVD.---}
Equations~(\ref{eq:kfermion}) and (\ref{eq:kboson}) show that particle statistics act as a change of metric in frequency space.  Define weighted inner products
\begin{equation}
\langle f,g\rangle_\nu=\int_{-1}^{1}f(y)g(y)\rho^\nu(y)\,\dd y,
\end{equation}
where
\begin{equation}
\rho^F=\cosh^2(cy),\quad
\rho^B=\frac{\sinh^2(cy)}{y^2},
\label{eq:metric}
\end{equation}
with the regular limit understood at $y=0$ for bosons.  Combining Eq.~(\ref{eq:slepian-expansion}) with Eq.~(\ref{eq:metric}) gives the generalized SVD
\begin{equation}
K^\nu(x,y)=\sum_{n=0}^{\infty}S^\nu_n U_n(x)V^\nu_n(y),
\label{eq:svd}
\end{equation}
with
\begin{align}
U_n(x)&=\psi_n(x;c),\\
V^F_n(y)&=\frac{{\rm sgn}(\mu_n)\psi_n(y;c)}{\cosh(cy)},&
S^F_n&=\frac{|\mu_n|}{2},\\
V^B_n(y)&=\frac{{\rm sgn}(\mu_n)y\psi_n(y;c)}{\sinh(cy)},&
S^B_n&=\frac{\wmax|\mu_n|}{2}.
\label{eq:svd-pieces}
\end{align}
Thus numerical kernel diagonalization in this weighted metric is replaced by the spectral algebra of a second-order differential operator; it is complementary to the standard $L^2$ IR SVD, whose effective rank at a fixed tolerance can be smaller after thermal weighting \cite{Misawa2026}.

The same gauge transformation gives the physical interpretation.  Writing $\psi_n(y;c)=g^\nu(y)V^\nu_n(y)$ with $g^F=\cosh(cy)$ and $g^B=\sinh(cy)/y$, the right singular functions obey a self-adjoint Sturm-Liouville equation,
\begin{equation}
\frac{1}{\rho^\nu}\frac{\dd}{\dd y}
\left[(1-y^2)\rho^\nu\frac{\dd V^\nu_n}{\dd y}\right]
 +{\cal V}^\nu(y)V^\nu_n
=\chi_n V^\nu_n .
\label{eq:gauge-sl}
\end{equation}
For fermions,
\begin{equation}
{\cal V}^F(y)=c^2-2cy\tanh(cy),
\label{eq:fermion-potential}
\end{equation}
up to the irrelevant constant shift fixed by the convention for $\chi_n$.  The thermal bath therefore appears as a gauge field that reshapes the frequency-space measure and suppresses statistically inaccessible high-energy information.  Fermions and bosons differ only through this metric and its associated effective potential; the singular-value cliff itself is governed by the universal Slepian channel.  The orthogonality relations and the gauge-transformed Sturm-Liouville equations are derived in Sec.~S3 of the Supplemental Material.

{\it Optimal sampling from a colleague matrix.---}
The analytical basis also fixes where imaginary-time data should be measured.  Truncating at $N=\lceil N_c\rceil$, the extended Gaussian quadrature nodes are the zeros of the boundary basis function,
\begin{equation}
\psi_N(x_i;c)=0,\qquad i=1,\ldots,N,
\label{eq:nodes}
\end{equation}
so uniform sampling is not information-theoretically optimal.  These nodes can be obtained without Newton iterations or prior function evaluations.  Expand
\begin{equation}
\psi_N(x;c)\simeq \sum_{m=0}^{M}d_mP_m(x),
\label{eq:legendre}
\end{equation}
where $P_m$ are Legendre polynomials and the coefficients $d_m$ are already produced by the parity-separated tridiagonal eigenproblem associated with Eq.~(\ref{eq:commutation}).  Using
\begin{equation}
xP_m(x)=\alpha_mP_{m+1}(x)+\gamma_mP_{m-1}(x),
\end{equation}
where
\begin{equation}
\alpha_m=\frac{m+1}{2m+1},\quad
\gamma_m=\frac{m}{2m+1},
\end{equation}
these coefficients form the Jacobi matrix $J$ for multiplication by $x$ on
$\bm P=(P_0,\ldots,P_{M-1})^T$.  At a zero of Eq.~(\ref{eq:legendre}),
$P_M(x_i)=-d_M^{-1}\sum_{m=0}^{M-1}d_mP_m(x_i)$.  Eliminating $P_M$ closes the recurrence into the colleague eigenproblem
\begin{equation}
C\bm P(x_i)=x_i\bm P(x_i),
\end{equation}
where
\begin{equation}
C=J-\frac{\alpha_{M-1}}{d_M}\bm e_{M-1}
(d_0,\ldots,d_{M-1}).
\label{eq:colleague}
\end{equation}
The physical nodes are the $N$ eigenvalues of $C$ in $[-1,1]$, and their quadrature weights follow from the corresponding eigenvectors by the usual Golub-Welsch moment construction \cite{GolubWelsch1969,XiaoRokhlinYarvin2001}.  The colleague-matrix formulation \cite{Good1961,NoferiniPerez2017} replaces heuristic root searches by a small deterministic linear-algebra problem whose dimension is set by $N_c$ plus a modest Legendre cutoff margin; implementation details are summarized in Sec.~S6 of the Supplemental Material.

\begin{figure}[t]
\includegraphics[width=\columnwidth]{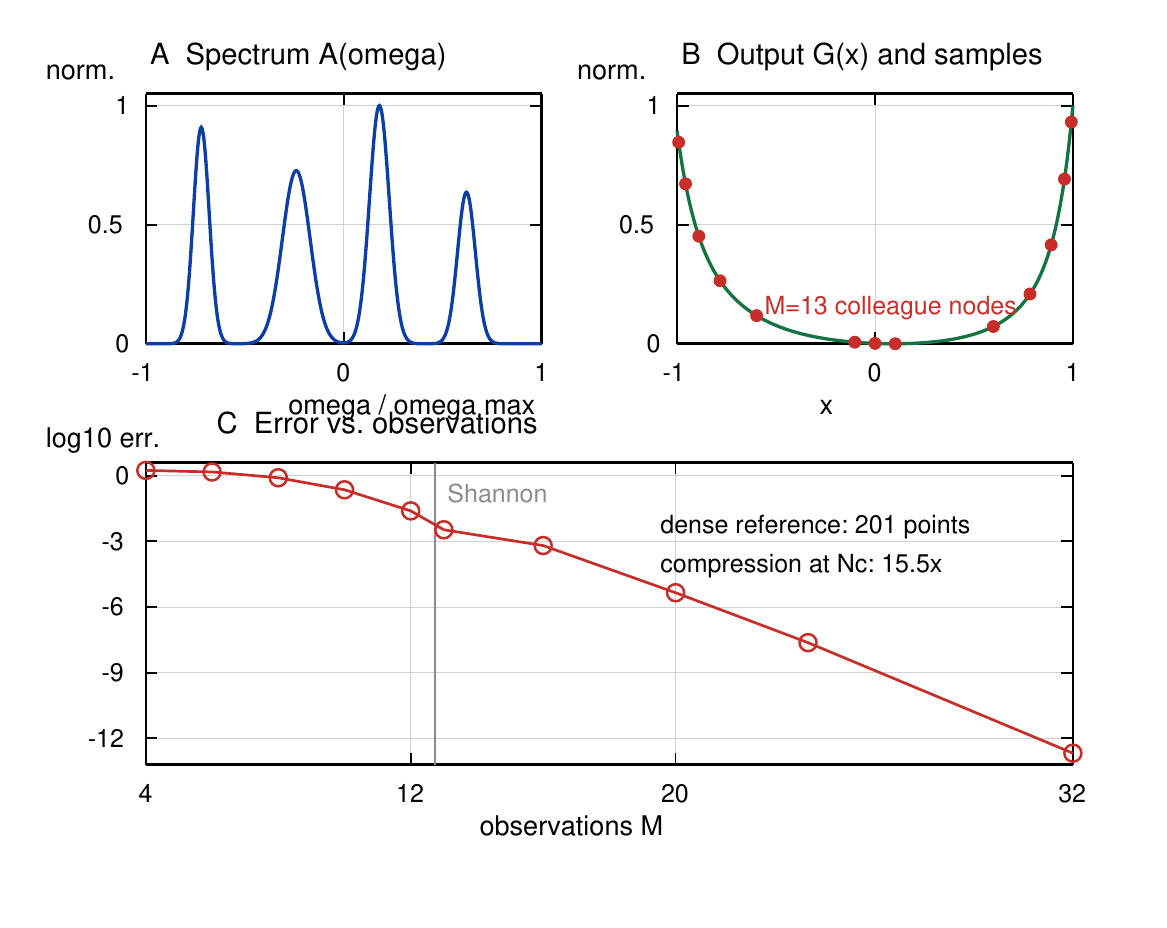}
\caption{Numerical sampling experiment; see Sec.~S6 of the Supplemental Material for algorithmic details.  (A) Original spectrum $A(\omega)$ made of four Gaussian peaks, shown as a function of $y=\omega/\wmax$.  (B) Imaginary-time output obtained from the fermionic kernel; red points are the $M=\lceil N_c\rceil=13$ colleague-matrix observation nodes.  (C) Relative reconstruction error of $G(x)$ as the number of observations $M$ is varied.  The sharp improvement at the Shannon scale shows that the compressed grid preserves the information transmitted by the kernel.}
\label{fig:sampling-demo}
\end{figure}

{\it Numerical sampling test.---}
We finally illustrate the sampling step in a controlled forward problem.  A model spectrum $A(\omega)$ consisting of four Gaussian peaks, plotted against $y=\omega/\wmax$, is transformed by the fermionic kernel with $c=20$, or $\beta\wmax=40$, for which $N_c=\beta\wmax/\pi=12.73$.  The dense reference uses $N_{\rm dense}=201$ uniformly spaced imaginary-time points.  Keeping only $M$ colleague-matrix observation nodes, Fig.~\ref{fig:sampling-demo} shows that the error is large when $M<N_c$, but drops to $3.4\times10^{-3}$ at $M=13$ and to $6.4\times10^{-4}$ at $M=16$, corresponding to a $201/13=15.5$ compression factor at the Shannon scale.  This should be distinguished from the final reconstruction accuracy of $A(\omega)$: the sparse-modeling method already treats the ill-conditioned directions in an information-theoretically sound way by keeping $S_n$ in the forward matrix instead of dividing by it \cite{Otsuki2017}.  The Slepian construction therefore does not evade the Shannon limit; it identifies the pure-channel limit analytically and converts it into a deterministic sampling prescription.  Existing uniform data can still be projected onto the analytical $U_n,S_n$ and processed by the same ADMM scheme, as checked in Sec.~S7 of the Supplemental Material.

{\it Conclusion and outlook.---}
Analytic continuation is difficult because thermal imaginary-time evolution is an information channel with exponentially small singular directions.  By separating the universal Laplace transmission factor from the statistical metric, we have shown that this channel is analytically controlled by Slepian's time-bandwidth theory.  Fermionic and bosonic kernels acquire their physical distinctions through gauge-transformed measures and effective potentials, while the Shannon number $N_c=\beta\wmax/\pi$ bounds the capacity of the pure Laplace channel.  This perspective also clarifies why singular-value-embedded ADMM is effective: it regularizes the information that survives the kernel rather than trying to recover modes that are not transmitted.  The new practical gain is the colleague-matrix recipe for the optimal QMC sampling grid.  It provides a mathematical basis for true compressed sensing in imaginary time: measure only the coordinates that span the transmitted information, and leave the exponentially unstable complement unmeasured.

{\it Acknowledgment.---}
We received financial supports by Cross-ministerial Strategic Innovation Promotion Program (SIP) from the Cabinet Office (No. 23836436).

\bibliography{references}

@article{JarrellGubernatis1996,
  author = {Jarrell, Mark and Gubernatis, J. E.},
  title = {Bayesian inference and the analytic continuation of imaginary-time quantum Monte Carlo data},
  journal = {Physics Reports},
  volume = {269},
  number = {3},
  pages = {133--195},
  year = {1996},
  doi = {10.1016/0370-1573(95)00074-7}
}

@article{VidbergSerene1977,
  author = {Vidberg, H. J. and Serene, J. W.},
  title = {Solving the {Eliashberg} equations by means of {$N$}-point {Pad{\'e}} approximants},
  journal = {Journal of Low Temperature Physics},
  volume = {29},
  number = {3-4},
  pages = {179--192},
  year = {1977},
  doi = {10.1007/BF00655090}
}

@article{SilverSiviaGubernatis1990,
  author = {Silver, R. N. and Sivia, D. S. and Gubernatis, J. E.},
  title = {Maximum-entropy method for analytic continuation of quantum {Monte Carlo} data},
  journal = {Physical Review B},
  volume = {41},
  number = {4},
  pages = {2380--2389},
  year = {1990},
  doi = {10.1103/PhysRevB.41.2380}
}

@article{Sandvik1998,
  author = {Sandvik, Anders W.},
  title = {Stochastic method for analytic continuation of quantum {Monte Carlo} data},
  journal = {Physical Review B},
  volume = {57},
  number = {17},
  pages = {10287--10290},
  year = {1998},
  doi = {10.1103/PhysRevB.57.10287}
}

@article{ShaoSandvik2023,
  author = {Shao, Hui and Sandvik, Anders W.},
  title = {Progress on stochastic analytic continuation of quantum {Monte Carlo} data},
  journal = {Physics Reports},
  volume = {1003},
  pages = {1--88},
  year = {2023},
  doi = {10.1016/j.physrep.2022.11.002}
}

@article{Otsuki2017,
  author = {Otsuki, Junya and Ohzeki, Masayuki and Shinaoka, Hiroshi and Yoshimi, Kazuyoshi},
  title = {Sparse modeling approach to analytical continuation of imaginary-time quantum Monte Carlo data},
  journal = {Physical Review E},
  volume = {95},
  number = {6},
  pages = {061302},
  year = {2017},
  doi = {10.1103/PhysRevE.95.061302}
}

@article{Yoshimi2019,
  author = {Yoshimi, Kazuyoshi and Otsuki, Junya and Motoyama, Yuichi and Ohzeki, Masayuki and Shinaoka, Hiroshi},
  title = {{SpM}: Sparse modeling tool for analytic continuation of imaginary-time Green's function},
  journal = {Computer Physics Communications},
  volume = {244},
  pages = {319--323},
  year = {2019},
  doi = {10.1016/j.cpc.2019.07.001}
}

@article{Shinaoka2017,
  author = {Shinaoka, Hiroshi and Otsuki, Junya and Ohzeki, Masayuki and Yoshimi, Kazuyoshi},
  title = {Compressing Green's function using intermediate representation between imaginary-time and real-frequency domains},
  journal = {Physical Review B},
  volume = {96},
  number = {3},
  pages = {035147},
  year = {2017},
  doi = {10.1103/PhysRevB.96.035147}
}

@article{Li2020,
  author = {Li, Jia and Wallerberger, Markus and Chikano, Naoya and Yeh, Chia-Nan and Gull, Emanuel and Shinaoka, Hiroshi},
  title = {Sparse sampling approach to efficient ab initio calculations at finite temperature},
  journal = {Physical Review B},
  volume = {101},
  number = {3},
  pages = {035144},
  year = {2020},
  doi = {10.1103/PhysRevB.101.035144}
}

@article{Shinaoka2020,
  author = {Shinaoka, Hiroshi and Geffroy, Dominique and Wallerberger, Markus and Otsuki, Junya and Yoshimi, Kazuyoshi and Gull, Emanuel and Kune{\v{s}}, Jan},
  title = {Sparse sampling and tensor network representation of two-particle Green's functions},
  journal = {SciPost Physics},
  volume = {8},
  pages = {012},
  year = {2020},
  doi = {10.21468/SciPostPhys.8.1.012}
}

@article{Kaye2022,
  author = {Kaye, Jason and Chen, Kun and Parcollet, Olivier},
  title = {Discrete {Lehmann} representation of imaginary time Green's functions},
  journal = {Physical Review B},
  volume = {105},
  number = {23},
  pages = {235115},
  year = {2022},
  doi = {10.1103/PhysRevB.105.235115}
}

@article{Motoyama2022,
  author = {Motoyama, Yuichi and Yoshimi, Kazuyoshi and Otsuki, Junya},
  title = {Robust analytic continuation combining the advantages of the sparse modeling approach and the {Pad{\'e}} approximation},
  journal = {Physical Review B},
  volume = {105},
  number = {3},
  pages = {035139},
  year = {2022},
  doi = {10.1103/PhysRevB.105.035139}
}

@article{FeiYehGull2021,
  author = {Fei, Jiani and Yeh, Chia-Nan and Gull, Emanuel},
  title = {{Nevanlinna} Analytical Continuation},
  journal = {Physical Review Letters},
  volume = {126},
  number = {5},
  pages = {056402},
  year = {2021},
  doi = {10.1103/PhysRevLett.126.056402}
}

@article{HuangYang2022,
  author = {Huang, Dongchen and Yang, Yi-feng},
  title = {Learned optimizers for analytic continuation},
  journal = {Physical Review B},
  volume = {105},
  number = {7},
  pages = {075112},
  year = {2022},
  doi = {10.1103/PhysRevB.105.075112}
}

@article{HuangACFlow2026,
  author = {Huang, Li},
  title = {{ACFlow} 2.0: An open source toolkit for analytic continuation of quantum {Monte Carlo} data},
  journal = {Computer Physics Communications},
  volume = {321},
  pages = {110038},
  year = {2026},
  doi = {10.1016/j.cpc.2026.110038}
}

@article{Misawa2026,
  author = {Misawa, Takahiro},
  title = {Analytic Origin of Green-Function Compression in the Intermediate Representation},
  journal = {arXiv preprint},
  year = {2026},
  eprint = {2605.24814},
  archivePrefix = {arXiv},
  primaryClass = {cond-mat.str-el},
  doi = {10.48550/arXiv.2605.24814}
}

@article{SlepianPollak1961,
  author = {Slepian, D. and Pollak, H. O.},
  title = {Prolate spheroidal wave functions, Fourier analysis and uncertainty. I},
  journal = {Bell System Technical Journal},
  volume = {40},
  number = {1},
  pages = {43--63},
  year = {1961},
  doi = {10.1002/j.1538-7305.1961.tb03976.x}
}

@article{LandauPollak1961,
  author = {Landau, H. J. and Pollak, H. O.},
  title = {Prolate spheroidal wave functions, Fourier analysis and uncertainty. II},
  journal = {Bell System Technical Journal},
  volume = {40},
  number = {1},
  pages = {65--84},
  year = {1961},
  doi = {10.1002/j.1538-7305.1961.tb03977.x}
}

@article{LandauPollak1962,
  author = {Landau, H. J. and Pollak, H. O.},
  title = {Prolate spheroidal wave functions, Fourier analysis and uncertainty. III: The dimension of the space of essentially time- and band-limited signals},
  journal = {Bell System Technical Journal},
  volume = {41},
  number = {4},
  pages = {1295--1336},
  year = {1962},
  doi = {10.1002/j.1538-7305.1962.tb03279.x}
}

@article{Slepian1978,
  author = {Slepian, D.},
  title = {Prolate spheroidal wave functions, Fourier analysis and uncertainty. V: The discrete case},
  journal = {Bell System Technical Journal},
  volume = {57},
  number = {5},
  pages = {1371--1430},
  year = {1978},
  doi = {10.1002/j.1538-7305.1978.tb02104.x}
}

@article{Good1961,
  author = {Good, I. J.},
  title = {The colleague matrix, a {Chebyshev} analogue of the companion matrix},
  journal = {Quarterly Journal of Mathematics},
  volume = {12},
  number = {1},
  pages = {61--68},
  year = {1961},
  doi = {10.1093/qmath/12.1.61}
}

@article{GolubWelsch1969,
  author = {Golub, Gene H. and Welsch, John H.},
  title = {Calculation of Gauss quadrature rules},
  journal = {Mathematics of Computation},
  volume = {23},
  number = {106},
  pages = {221--230},
  year = {1969},
  doi = {10.1090/S0025-5718-69-99647-1}
}

@article{XiaoRokhlinYarvin2001,
  author = {Xiao, Hong and Rokhlin, Vladimir and Yarvin, Norman},
  title = {Prolate spheroidal wavefunctions, quadrature and interpolation},
  journal = {Inverse Problems},
  volume = {17},
  number = {4},
  pages = {805--838},
  year = {2001},
  doi = {10.1088/0266-5611/17/4/315}
}

@article{NoferiniPerez2017,
  author = {Noferini, Vanni and P{\'e}rez, Javier},
  title = {Chebyshev rootfinding via computing eigenvalues of colleague matrices: When is it stable?},
  journal = {Mathematics of Computation},
  volume = {86},
  number = {306},
  pages = {1741--1767},
  year = {2017},
  doi = {10.1090/mcom/3149}
}

\end{document}